# A full free spectral range tuning of p-i-n doped Gallium Nitride microdisk cavity

Nan Niu[1], Tsung-Li Liu[1], Igor Aharonovich[1], Kasey J Russell[1], Alexander Woolf[1], Thomas C. Sadler[2], Haitham A.R. El-Ella[2], Menno J. Kappers[2], Rachel A Oliver[2], and Evelyn L. Hu[1]

[1]School of Engineering and Applied Sciences, Harvard University, Cambridge, MA, 02138, USA

[2]Department of Materials Science and Metallurgy, University of Cambridge, Pembroke Street, Cambridge CB2 3QZ, United Kingdom

Email: nanniu@fas.harvard.edu

**Abstract**

Effective, permanent tuning of the whispering gallery modes (WGMs) of p-i-n doped GaN microdisk cavity with embedded InGaN quantum dots over one free spectral range is successfully demonstrated by irradiating the microdisks with a ultraviolet laser (380nm) in DI water. For incident laser powers between 150 and 960 nW, the tuning rate varies linearly. Etching of the top surface of the cavity is proposed as the driving force for the observed shift in WGMs, and is supported by experiments. The tuning for GaN/InGaN microdisk cavities is an important step for deterministically realizing novel nanophotonic devices for studying cavity quantum electrodynamics.

In recent years, there has been tremendous progress in the understanding of light-matter interactions (cavity quantum electrodynamics, or cQED) in semiconductor systems[1-8]. Strong coupling has been observed for GaAs-based cavities coupled to embedded InGaAs quantum dots[2, 3, 4], and studies have also extended to other semiconductor materials[5]. Such studies provide a promising route to the realization of quantum information technology[6-9] and ultra-efficient light emitting devices[10, 11]. In particular, InGaN quantum dots (QDs), well coupled to GaN-based optical cavities offer the potential of highly efficient devices operating at room temperature in the visible to UV wavelengths[12-15]. An outstanding challenge in achieving coupled QD-cavity systems is the need to realize both spatial and frequency resonance of QD to cavity modes. This work describes a well-controlled technique for tuning the whispering gallery modes (WGMs) of an InGaN/GaN microdisk. The technique achieves a shift of the cavity modes over one free spectral range without degradation of the quality factor (Q) of the cavity. This work builds on our earlier observation of a selective, sensitive tuning mechanism[16], but makes several important advances: (1) Tuning was carried out under low incident laser powers to achieve a well-controlled, linear tuning rate. While only small degradations in Q were observed in our earlier work, the tuning achieved in this work occurred *without* degradation of the Q of the microdisk cavity. This was despite the fact that modes were tuned over 20 nanometer shifts in wavelength. (2) Despite the lower powers used in these experiments, the tuning rates achieved (in wavelength shift/incident power/time) were substantially higher than previously achieved. We believe that this is because the material that forms the microdisks in these experiments is a p-i-n structure, compared to the previously non-intentionally doped (nid) structures[16]. The different built-in electric field in these structures produces a more uniform, faster and well-controlled etch of the structure. (3) Through selective masking of the microdisk, we have experimentally verified that the tuning mechanism is related to a photo-enhanced etch of the microdisk.

The material structure was grown by metalorganic vapor phase epitaxy on a c-plane GaN/Al$_2$O$_3$ pseudosubstrate[17, 18]. A sacrificial superlattice of *n*-doped In$_{0.07}$Ga$_{0.93}$N/In$_{0.05}$Ga$_{0.95}$N (200 nm total depth) was first formed, capped by 30 nm of n-GaN. 20 nm of AlGaN formed a thin layer to enhance the robustness of the PEC process, followed by the growth of 10 nm of nid GaN on top of which the InGaN quantum dots were grown[19]. A 10 nm GaN capping layer was then grown at the same temperature as the QDs, followed by a further 20 nm of nid GaN grown at 1000 °C. The growth of 30 nm of p-GaN completed the material structure. (The growth and anneal conditions for the p-GaN have been previously demonstrated not to damage the QDs[20], and have also been shown to have no measureable effect on the sacrificial superlattice for the compositions used here).

Fabrication of the microdisks was carried out by first dispersing 2 μm diameter SiO$_2$ beads onto the substrate surface. The beads served as masks for the subsequent inductively coupled plasma (ICP) etch of the material in 25 sccm of Cl$_2$ and Argon gas for an approximate depth of 520 nm. The SiO$_2$ beads were subsequently removed, and photoelectrochemical (PEC) etching in 0.004 M HCl was then performed to selectively etch the In$_{0.07}$Ga$_{0.93}$N/In$_{0.05}$Ga$_{0.95}$N superlattice, forming the final microdisk structures. Details of the full PEC process can be found elsewhere[21]. Figure 1(a) shows a scanning electron microscope (SEM) image of the microdisk. The InGaN QD active layer is located in the middle of the disk membrane sandwiched by the p- and n-doped GaN layers as shown in Figure 1(b). The diameter of the disks is ~ 2 μm and the thickness of the disk membrane is ~ 120 nm. Optical characterization of the microdisks was performed using a frequency-doubled titanium sapphire laser emitting at 380 nm, an energy below the bandgap of GaN, through a long working distance objective (×100, numerical aperture (NA)=0.5). The beam diameter is ~500 nm. The emission from the microdisks was collected through the same objective and directed into a spectrometer for analysis. Figure 1(c) shows a room temperature photoluminescence (PL) spectrum from the microdisk. Distinct modes can be seen decorating the broad background emission of quantum dots. The Qs of the first order WGMs are ~ 1500.

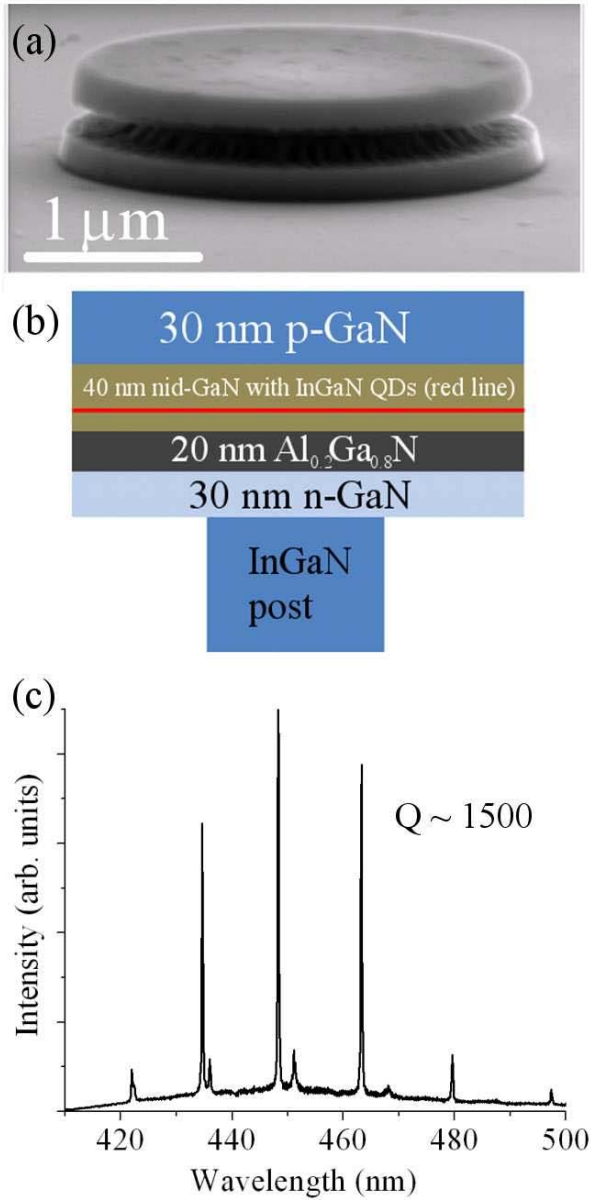

Fig. 1 (a) SEM image of the InGaN/GaN microdisk. (b) The material structure of microdisk. The red line in the middle of the disk membrane represents the InGaN QDs layer. (c) PL spectrum recorded from the disk showing the WGMs with Q approximately 1500.

In order to tune the WGMs, the disks were immersed in water within a small cell and subjected to illumination with laser beam directly incident onto the p-GaN of the disk structure. The description of the cell is given previously[16]. After illumination in water for a given incident power and time, the microdisks were removed from the cell and dried. A spectrum of the microdisk was then taken in air to more precisely determine the shift in the mode. Moreover, we have ascertained that direct illumination of microdisks in air does not induce a mode shift. Figure 2(a) details the shift of a given mode for 60 seconds illumination at incident powers of 150 nW, 450 nW, and 960 nW, respectively, all measured directly underneath the focusing objective. Figure 2(b) shows the shift in a mode as a function of time, measured at a constant laser power of 450nW. In both cases, the linear shift in mode wavelength with time, at constant power, or with power at a constant time, provides the possibility of deterministic tuning of one WGM into resonance with a quantum dot transition. Figure 2(c) is a histogram of 'normalized' tuning rate in nanometers (of wavelength shift)/incident power/time of each of 15 different tested microdisks. The standard deviation of the rates falls within 13% of the rate value. This result shows homogeneity in the tuning process and implies that the material properties among different locations of the sample are relatively uniform.

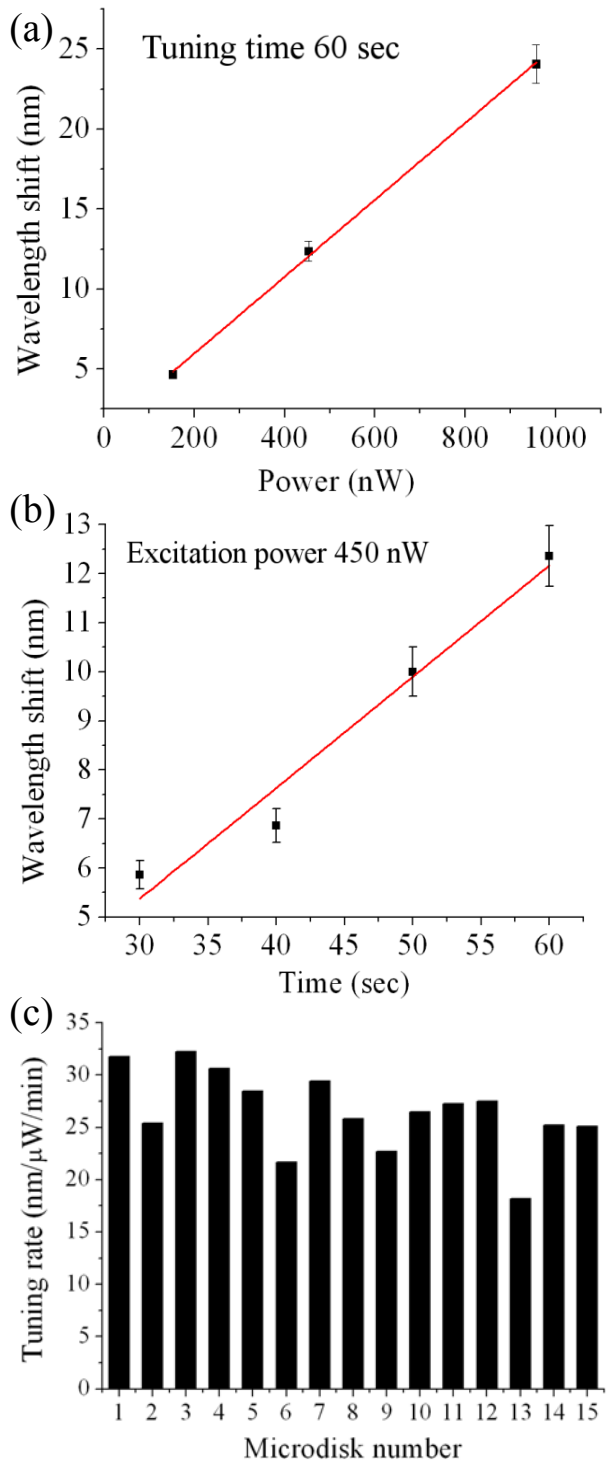

Fig. 2 (a) A plot showing tuning range vs. different excitation powers with constant tuning time of 60 seconds. (b) Plot showing tuning range vs. time with constant power of 450 nW. (c) A histogram of normalized tuning rates (nm/µW/min) of 15 microdisks.

To further demonstrate the precise control of WGM shifts possible with this technique, we attempted full spectral range tuning on several microdisks. A chosen microdisk was tuned in 5 consecutive 10-second intervals. After each tuning cycle, the sample was taken out of water, dried and a spectrum was taken in air. The results are displayed in Figure 3. The spectra are offset in the y-direction, for clarity, and the bottom spectrum is the PL signature of the disk before tuning is commenced. The subsequent spectra represent the changes after successive additional 10 seconds of tuning. The shadowed box highlights the evolution of a selected WGM. As the arrow indicates, this mode was tuned from its original wavelength to the emission line of the initial adjacent mode. No Q degradation is observed.

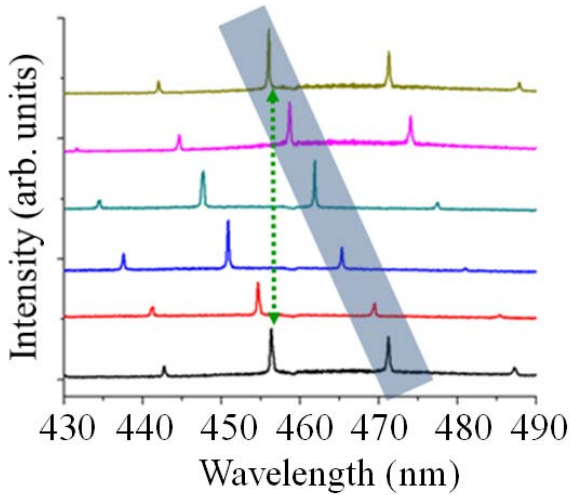

Fig. 3 PL spectrums of microdisk cavity over 50 seconds of tuning under power ~450nW with 380nm laser. The bottom spectrum is taken before tuning. The arrow indicates the evidence of tuning one WGM over one free spectral range. The shadow box indicates the evolution of the WGM. The times interval before each spectrum is 10s increasing from bottom to top.

In our previous work, we tentatively identified the tuning mechanism as a process similar to that seen in PEC etching of GaN[16]. This is a process in which photo-generated holes enhance the oxidation of the GaN, with the oxide subsequently dissolved in the electrolyte (water), leading to a slight etching of the material[21]. In this work, we carried out an experiment to directly validate this model, by masking a portion of the microdisk to mark changes in its geometry during the water-based tuning process. A masking structure 80 nm in diameter was formed by electron beam lithography using Fox-16™ resist. The mask is designed to prevent photo-induced changes in the underlying material, and is placed at the center of the microdisk to avoid overlapping with the WGMs, as shown in Figure 4(a). The microdisk was then placed in the water cell and subjected to 450 nW illumination for 60 seconds. After the laser irradiation, the sample was removed from the cell and immersed in buffered oxide etch (BOE, 7:1 $NH_4F$:HF) for 1 minute to completely remove the mask. Atomic Force Microscopy (AFM) was then done on the microdisk to characterize the surface roughness. Figure 4(b) shows the AFM trace in the vicinity of the original mask. The masked region is ~8 nm higher than the surrounding surface, indicating that the top surface of the microdisk is etched during the tuning process.

PL of the microdisk after the tuning shows a 14.2 nm blue shift in one selected WGM. COMSOL simulations modeled the mode shift expected for the microdisk having an 8 nm reduction in vertical dimension for the disk membrane. The result of the simulation indicates a blue shift of ~13.9 nm in the selected WGM. The close correspondence between the simulation and experimental results suggests that only the top, p-type GaN is etched during this process; our final discussion centers on the detailed mechanism of the photo-induced etching and tuning in these structures.

We have described the tuning process to be similar to PEC etching of GaN; it is notable that in our experiments, the light is incident on the p-GaN material of the microdisk, and that generally p-type semiconductors are not etched in PEC processes[22]. However, our 380 nm laser excitation generates electron-hole pairs only in the lower bandgap quantum dot layer, not in the surrounding GaN. Under these conditions, as Tamboli et al. have shown[23], PEC etching can be used effectively to etch the p-GaN. Figure 4(c) shows a simplified bandstructure of a p-i-n junction representing the disk membrane. Electron-holes pairs are generated in the quantum dot layer. The natural band-bending of the structure and the rather shallow confinements of the electrons and holes enable the holes to both drift and diffuse to the p-GaN surface, enhancing oxidation and subsequent etching. The photo-generated electrons are similarly swept to the n-GaN side of the material. We believe that the generation of electron-hole pairs at 380 nm, confined to the quantum dot region, and the influence of the built-in electric field in effectively sweeping the holes to top surface underlies the rapid tunability, with substantial mode shifts and no degradation of quality factor. By contrast, under illumination of 360 nm radiation at only 120 nW power for only 10 seconds, the mode shift observed is about 6 nm, but the Q is degraded from 1200 to 450. In this case, electron-hole pairs are generated throughout the volume of the microdisk, it appears that the underlying n-GaN is preferentially etched, but the structural symmetry of the disk is destroyed and there is a degradation of Q. In comparing the tuning of these p-i-n structures with the nid structures we had earlier characterized, we note that we see tuning of the modes at substantially lower powers (200 nW to 1 µW, for either 360 nm or 380 nm incident wavelength) than before (few hundred µW to ~ 1 nW). Illumination of the nid structures at 50 µW power and 380 nm produced no tuning at all. While the histogram of Figure 2(c) shows average tuning rates of ~20 nm/µW/min for the p-i-n microdisks, the tuning rates for the nid microdisks were 3 to 4 orders of magnitude less. Thus the bandstructure of the cavity material, relative to the excitation wavelength plays an important role in determining the predictability and control of the tuning process.

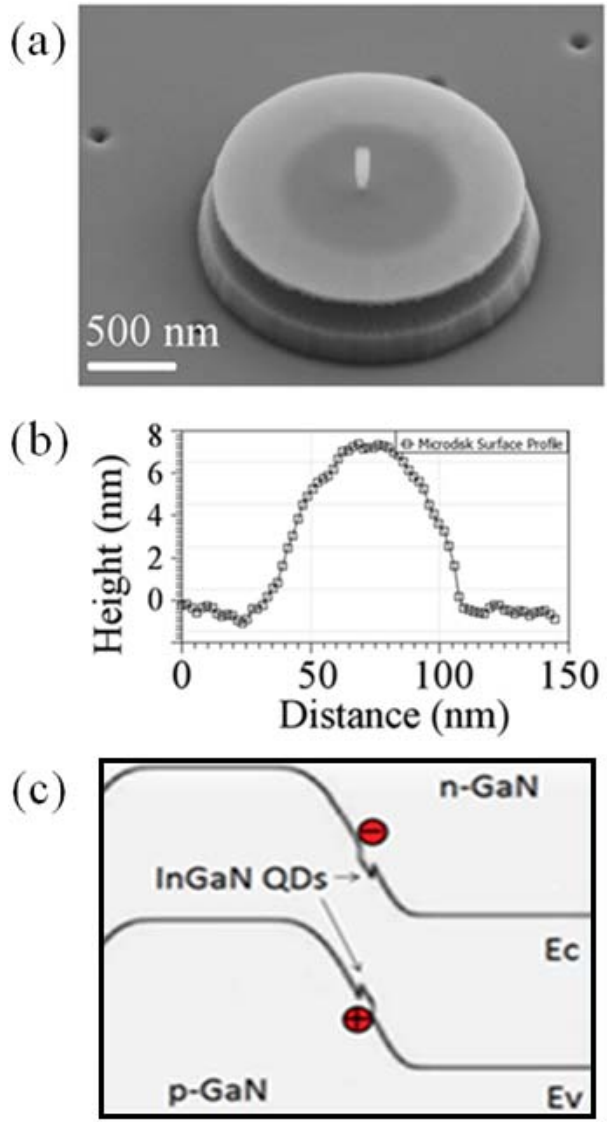

Fig. 4 (a) SEM image of the InGaN/GaN $\mu$–disk with FOx$^{TM}$ resist mask. (b) AFM scan of microdisk top surface profile in the vicinity of the removed resist mask. (c) A simplified bandstructure of a p-i-n junction of disk membrane. The quantum wells placed in middle represent the quantum dot layer. The red dots with plus and minus sign represent hole and electron excited by laser, respectively.

In conclusion, we have successfully demonstrated an effective, permanent tuning technique of the WGMs of a p-i-n doped GaN/InGaN microdisk cavity over one free spectral range without degrading their optical quality. It is found that the tuning rate is linearly dependent on the applied power of the laser excitation. Fine tuning can therefore be achieved and controlled by tuning the cavity at low incident laser powers.

We have confirmed the change in geometry brought about by the tuning process. Our technique constitutes an important step in achieving spectral resonances of WGMs with embedded InGaN QDs which is crucial for studying cavity QED and realizing novel optoelectronic devices.

**Acknowledgement**

The authors thank Dr. Andrew Magyar and Jonathan Lee for useful discussions and Qimin Quan for help with the simulations. This work was enabled by facilities available at the Center for Nanoscale Systems(CNS), a member of the National Nanotechnology Infrastructure Network (NNIN), which is supported by the National Science Foundation under NSF award no. ECS-0335765. CNS is part of the Faculty of Arts and Sciences at Harvard University. This work was also supported in part by the NSF Materials World Network (Award No. 1008480), the Engineering and Physical Sciences Research Council (Award No. EP/H047816/1), and the Royal Academy of Engineering. R.A. Oliver would like to acknowledge funding from the Royal Society.